\documentclass[english,a4paper]{article}
\usepackage{dcolumn}
\usepackage{bm}
\usepackage{graphicx}
\usepackage{amsmath}
\usepackage{amsfonts}
\usepackage{amssymb}
\usepackage{amsthm}
\usepackage{amstext}
\usepackage{amsbsy}
\usepackage{amsopn}
\usepackage{amscd}
\usepackage{amsxtra}

\def\openone{\leavevmode\hbox{\small1\kern-3.3pt\normalsize1}}

\usepackage[usenames,dvipsnames]{color}

\begin{document}

\title{Shaping of the time-evolution of field-free molecular orientation by THz laser pulses}
\author{R. Tehini\footnote{Lebanese University, EC2M Laboratory, Plateforme doctorale des Nanotechnologies, Faculty of Sciences II Fanar, Beirut,  Lebanon}, K. Hamraoui, D. Sugny\footnote{Laboratoire Interdisciplinaire Carnot de
Bourgogne (ICB), UMR 6303 CNRS-Universit\'e Bourgogne-Franche Comt\'e, 9 Av. A.
Savary, BP 47 870, F-21078 Dijon Cedex, France and Institute for Advanced Study, Technische Universit\"at M\"unchen, Lichtenbergstrasse 2 a, D-85748 Garching, Germany, dominique.sugny@u-bourgogne.fr}}

\maketitle

\begin{abstract}
We present a theoretical study of the shaping of the time-evolution of field-free orientation of linear molecules. We show the extend to which the degree of orientation can be steered along a desired periodic time-dependent signal. The objective of this study is not to optimize molecular orientation but to propose a general procedure to precisely control rotational dynamics through the first moment of the molecular axis distribution. Rectangular and triangular signals are taken as illustrative examples. At zero temperature, we compute the quantum states leading to such field-free dynamics. A TeraHertz laser pulse is designed to reach these states by using optimal control techniques. The investigation is extended to the case of non-zero temperature. Due to the complexity of the dynamics, the control protocol is derived with a Monte-Carlo simulated annealing algorithm. A figure of merit based on the Fourier coefficients of the degree of orientation is used. We study the robustness of the control process against temperature effects and amplitude variations of the electric field.
\end{abstract}

\section{Introduction}\label{sec1}
The goal of quantum control is to design external electromagnetic pulses for realizing different tasks such as population transfer between different quantum states or quantum gates~\cite{Glaser:15,Brif:10}. In atomic and molecular physics, this domain has many applications extending from photochemistry to quantum computation~\cite{Glaser:15,Brif:10,Shapiro:12,Warren:93,Shapiro:03}. Molecular alignment and orientation are well-established topics in quantum control from both the experimental and theoretical points of view~\cite{Stapelfeldt:03,Seideman:05,RMP,lemeshko:2013}. The control of the alignment process is by now well understood in the adiabatic or sudden regime~\cite{Friedrich:95,Stapelfeldt:03,Seideman:05,Leibscher:03,Leibscher:04,Peronne:03,Salomon:05,Velotta:01}. Recent works have shown the possibility to extend the standard control framework by considering, e.g., the deflection of aligned molecules~\cite{Gershnabel:10} and the role of collisional effects~\cite{RamakrishnaSeideman2005,Viellard:08,Viellard:13}. The shaping of fied-free alignment dynamics has also been extensively investigated with studies showing, to mention a few, the planar alignment~\cite{Hoque:11}, the unidirectional rotation of molecular axis~\cite{korech:2013,Steinitz:14,karras:2015}, alignment alternation~\cite{daems:2005} or the control of rotational wave packet dynamics~\cite{spanner:2004}. On the theoretical side, several control mechanisms and protocols have been proposed~\cite{Averbukh:01,Coudert:17,Daems:05,Dion:05,Lapert:12,Sugny:14,
Tehini:08,Tehini:12,Henriksen:99,Kitano:11,Li:13,Omiste:16,Ortigoso:12,Wu:10,Yoshida:14,spanner:2012,
shu:2013,liao:2013,Znakovskaya:2014,kurosaki:2014,qin:2014,trippel:2015,kallush:2015,takei:2016,damari:2017} to achieve molecular orientation. Some of them have been demonstrated
experimentally~\cite{Nelson:11,Babilotte:16,Ghafur:09,Goban:08,Znakovskaya:09,frumker:2012,frumker:2012b,dimitrovski:2011}, in particular in field-free conditions by using TeraHertz (THz) laser fields~\cite{Nelson:11,Babilotte:16}. However, the vast majority of control strategies developed so far has investigated the optimization of the degree of orientation at a given time.

We propose in this paper to explore another aspect of field-free molecular orientation, that is the shaping of the time-evolution of orientation dynamics by THz laser pulses. In the same direction, note that the tracking control of molecular orientation has been recently studied numerically~\cite{tracking:2018}. Using the fact that the time evolution of the orientation dynamics can be viewed as a truncated Fourier series, we show theoretically that the degree of orientation can be steered along a desired periodic signal with a zero time-integrated area. At zero temperature, target states are defined for periodic signals with non-zero Fourier coefficients. The role of the Gibbs phenomenon is also discussed. The rectangular and triangular waveforms are taken as illustrative examples. Using optimal control techniques, we design THz laser fields able to bring the system to the different target states. We extend this analysis to non-zero temperature. In this case, due to the complexity of the dynamics, target states cannot be uniquely defined. The control protocol is derived with a Monte-Carlo simulated annealing algorithm with a figure of merit based on the Fourier coefficients of the degree of orientation. Simple approximations of the control fields are achieved. Finally, we investigate the robustness of the shaping against temperature effects and variations of the control field. Possible experimental implementations are discussed.

The paper is organized as follows. The model system is presented in Sec.~\ref{sec2}. The shaping of the time-evolution of molecular orientation at zero temperature is investigated in Sec.~\ref{sec3} with the definition of the target states and the design of the corresponding control fields. An extension to nonzero temperature is proposed in Sec.~\ref{sec4}. Conclusions and prospective views are given in Sec.~\ref{sec5}.

\section{Model system}\label{sec2}

We consider the control of a linear polar molecule in its ground vibronic state by means of a linearly polarized (along the $z$- axis of the laboratory frame) THz laser field $E(t)$. Within the rigid rotor approximation, the dynamics of the system are governed by the following Hamiltonian~\cite{Stapelfeldt:03,Seideman:05}:
\begin{equation}\label{Hamiltonian}
H(t)=BJ^2-\mu_{0}E(t)\cos\theta,
\end{equation}
where $B$ is the rotational constant of the molecule, $\mu_{0}$ the permanent dipole moment, $\theta$ the polar angle between the direction of the polarization vector and the molecular axis and $J^{2}$ the angular momentum operator. We have verified that the effect of polarizability components can be neglected. The units used throughout the paper are atomic units unless otherwise specified. The Hilbert space associated with the dynamical system is spanned by the spherical harmonics $|j,m\rangle $, with $0\leq j$ and $-j\leq m\leq j$.  In this basis, The different operators have the following matrix elements~\cite{Zare:88,Varshalovich:88,BenHaj-Yedder:02}:
\begin{eqnarray*}\label{coefficients}
& & \langle j,m|J^2|j,m\rangle = j(j+1) \\
& & \langle j+1,m|\cos\theta|j,m\rangle =\frac{\sqrt{(j+1-m)(j+1+m)}}{\sqrt{(2j+1)(2j+3)}}=\alpha_{j,j+1}^{m}.
\end{eqnarray*}
At zero temperature, the initial state is $|\psi_0\rangle = |j_0=0,m_0=0\rangle$.  The interaction operator $\cos\theta$ does not couple the  wave functions with  different values of $m$, only the states $|j,m_0=0\rangle$ with $j\geq 0$ are populated by the laser excitation. For sake of simplicity, the coefficients $\alpha_{j,j+1}^{0}$ are denoted by $\alpha_{j,j+1}$ below. The degree of orientation is evaluated by the expectation value $\langle \cos\theta\rangle (t)=\langle \psi(t)|\cos\theta |\psi(t)\rangle$ where $|\psi(t)\rangle$ is the wave function of the system at time $t$. At a non-zero temperature,  different rotational states $|j_0,m_0\rangle$ are initially populated according to the Boltzmann distribution. The measure of orientation is the sum of the contributions coming from all of these states weighted by their respective population. This case is investigated in Sec.~\ref{sec4}. In the numerical simulations, the OCS and CO molecules are taken as illustrative examples. Numerical values of the molecular parameters are taken as $B=0.2059$~cm$^{-1}$, $\mu_0=0.712$~D for OCS and $B=1.92253$~cm$^{-1}$, $\mu_0=0.112$~D for CO.
\section{Shaping at zero temperature}\label{sec3}

\subsection{Description of the target states}
The molecule is subjected to a THz laser field in the interval $[0,t_0]$. When the electric field is switched off at $t=t_0$, the state of the system $|\psi_T\rangle$ can be expressed as $|\psi_T\rangle =\sum_{j=0}^{+\infty} C_{j} |j,0\rangle$, with the condition $\sum_{j=0}^{+\infty} |C_{j}|^2=1$.
The time evolution in field-free condition of $|\psi_T\rangle$ is given by:
\begin{equation}\label{timeevolpsi}
|\psi_T(t)\rangle = \sum_j C_je^{-iBj(j+1)(t-t_0)}|j,0\rangle .
\end{equation}
The degree of orientation can be written as follows:
\begin{equation}\label{cos}
\langle\cos\theta\rangle(t) =\sum_{j=0}^{+\infty}[\alpha_{j,j+1}C_{j+1}^*C_je^{2iB(j+1)(t-t_0)}+c.c.].
\end{equation}
Equation~\eqref{cos} can be interpreted as the Fourier expansion of $\langle\cos\theta\rangle(t)$. Introducing the time $\tau=t-t_0$ and the frequency $f_r=1/T_r$, with $T_r=\frac{\pi}{B}$ the rotational period, we arrive at:
\begin{equation}\label{cosFT}
\langle\cos\theta\rangle(\tau) =\sum_{j=0}^{+\infty}[K_j e^{i2\pi(j+1)f_r\tau}+c.c.],
\end{equation}
where $K_j=\alpha_{j,j+1}C_{j+1}^*C_j$. If you denote by $n=j+1$, we get the standard Fourier expansion of a periodic signal of period $T_r$ with a time-integrated zero area, since there is no zero frequency component in Eq.~\eqref{cosFT}.

We consider a generic signal defined by the complex coefficients of Eq.~\eqref{cosFT}: $K_j=|K_j|e^{i\psi_j}$, for $j=0,\cdots,j_{\textrm{max}}$. Using Eq.~\eqref{cos}, we arrive, for $j\geq 0$, at:
\begin{equation}
\begin{cases}
\alpha_{j,j+1}|C_{j+1}||C_j|=|K_j| \\
\phi_j-\phi_{j+1}=\psi_j,
\end{cases}
\end{equation}
where $\phi_{j}$ is the phase of the complex coefficient $C_{j}$. It is then straightforward to show that $|C_{j+1}|$ and $\phi_{j+1}$ can be expressed respectively in terms of $|C_{j}|$ and $\phi_{j}$ if the coefficients of the Fourier series $K_j$ are different from zero. We deduce that the time evolution of the degree of orientation can be shaped as any periodic signal with non-zero Fourier coefficients and with a zero time-integrated area. Two different types of functions, namely the rectangular and the triangular signals, are used as illustrative examples in Sec.~\ref{secex} to describe this method.
\subsection{Rectangular and triangular signals}\label{secex}
\begin{figure}[!ht]
	\centering
	\includegraphics[width=1\linewidth]{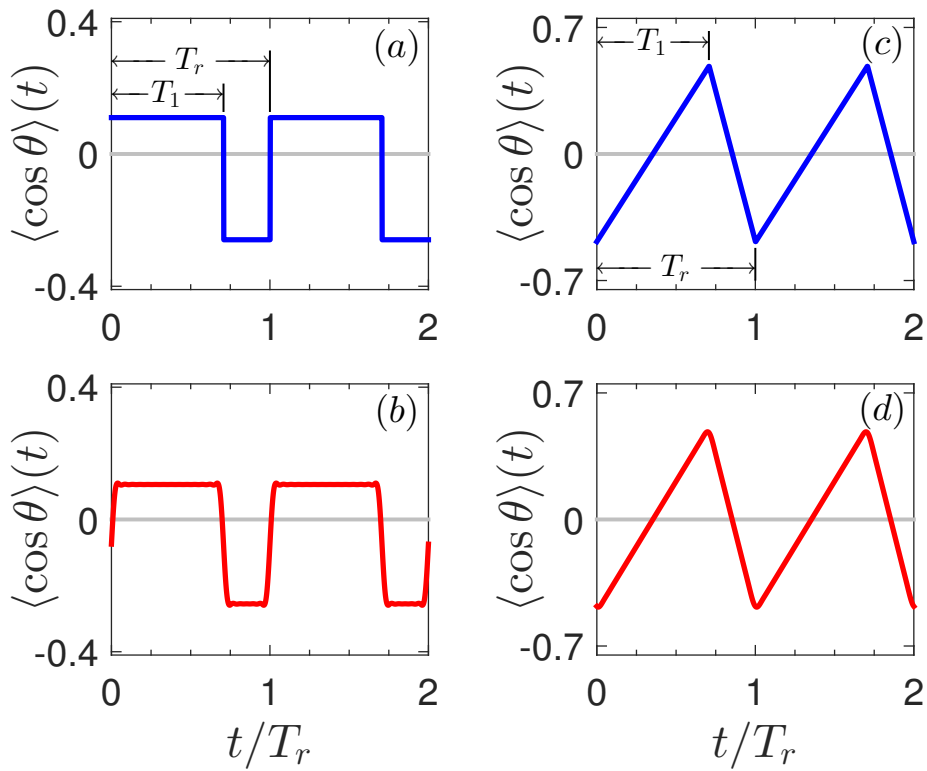}
	\caption{(Color online) Time evolution of the orientation of the OCS molecule at zero temperature for $j_{\max}=20$. Panels (a) and (c) represent the ideal rectangular and triangular signals, while panels (b) and (d) display the degree of orientation when the initial state is $|\psi_T\rangle$. The parameters $r$ and $A_0$ are respectively set to $r=1/\sqrt{2}$ and 1. The Gibbs-Wilbraham phenomenon has been corrected.}\label{fig1}
\end{figure}

We first consider the case of a rectangular signal. Its time evolution is displayed in Fig.~\ref{fig1}a. The zero area constraint leads to the relation:
\begin{equation}
A_{1}=-\frac{r}{1-r}A_{0},
\end{equation}
where $A_{0}$ and $A_{1}$ are respectively the maximum and the minimum amplitudes of the pulse over the period $T_{r}$ and $r$ the ratio $r=T_{1}/T_{r}$. For $0<t<T_r$, the signal $s(t)$ is defined as:
\begin{equation}\label{rectangular}
s(t)=
\begin{cases}
A_{0}, & 0<t<T_{1},\\
-\frac{r}{1-r}A_{0}, &  T_{1}<t<T_{r}.
\end{cases}
\end{equation}
Note that this expression depends on two free parameters $r\in ]0,1[$ and $A_0>0$.
The Fourier expansion of $s(t)$ is given by:
\begin{equation}\label{eqsquare}
\begin{aligned}
s(t) = \frac{1}{1 - r}
\sum_{n=1}^{\infty} \frac{A_{0}}{n\pi}e^{(i2\pi nf_r\tau-i\pi n r)} +c.c
\end{aligned}	
\end{equation}
The Fourier coefficients of Eq.~\eqref{eqsquare} can be directly identified with those of $\langle\cos\theta\rangle (\tau)$ in Eq.~\eqref{cosFT}. We obtain the following relations:
\begin{equation}
\begin{cases}
\alpha_{j,j+1}|C_{j+1}| |C_{j}|& =\frac{A_{0}}{\pi (j+1)(1 - r)} \sin[\pi (j+1)r],\\
\phi_{j}-\phi_{j+1}&=-\pi (j+1) r,
\end{cases}
\end{equation}
The condition of non-zero Fourier coefficients implies that $r$ is an irrational number. We set $\phi_0=0$ and we define $C_0$ so that the state $|\psi_T\rangle$ is normalized to 1. Note that a parameter $r$ larger than 0.5 allows us to obtain periodic signals with a maximum amplitude. In the numerical simulations, we reduce the physical Hilbert space to a finite subspace for which $j\leq j_{\textrm{max}}$. This reduction can be justified by the finite amount of energy that a laser field can transfer to the molecule. Another point to take into account is the Gibbs-Wilbraham phenomenon~\cite{Wilbraham:48,Gibbs:98,Hewitt:79}, which occurs at non-smooth points of the signal. To get rid of this artefact, we use a $\sigma$- approximation \cite{Lanczos:88,Hamming:87} which allows to smooth the truncated Fourier series. In this approximation, the different Fourier coefficients are multiplied by the factors $\sigma_{n}=\textrm{sinc}(n\pi/N)$, $n=1,\dots ,N$.

\begin{figure}[!ht]
	\centering
	\includegraphics[width=1\linewidth]{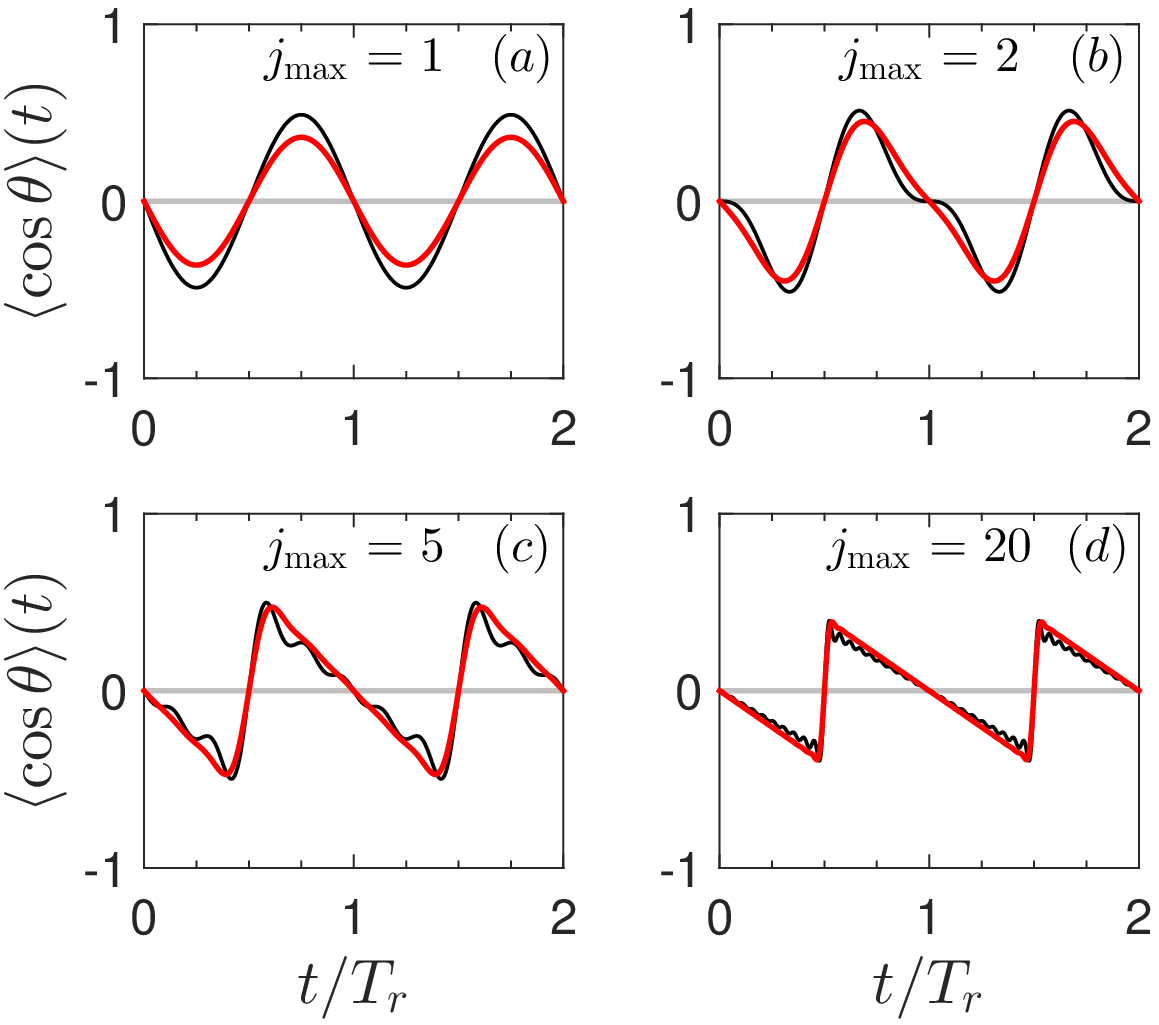}
	\caption{(Color online) Time evolution of the sawtooth shaped orientation of the OCS molecule for different values of $j_{\textrm{max}}$. The black and red (dark gray) solid lines represent the response without and with the correction to the Gibbs-Wilbraham phenomenon. The parameters $A_0$ and $r$ are set to 1 and $1/\sqrt{2}$.}
	\label{fig2}
\end{figure}

The same work can be done for a triangular signal as displayed in Fig.~\ref{fig1}c. The signal $s(t)$ can be expressed as:
\begin{equation}
s(t)=
\begin{cases}
A_0(\frac{2}{T_{1}}t-1), & 0<t<T_{1},\\
A_{0}(\frac{1+r}{1-r}-\frac{2}{T_{r}}\frac{1}{1-r}t), & T_{1}<t<T_{r},
\end{cases}
\end{equation}
where $A_0$ is the maximum amplitude and $r=T_1/T_r$. This function has the following Fourier expansion:
\begin{equation}
\begin{aligned}
u(t) =  &\sum_{n=1}^{\infty} \frac{A_{0}}{\pi^{2}n^{2}r(1-r)} \sin[\pi n r]\\
&\times e^{i[2\pi n f_r t - i \pi (n r+1/2)]}+c.c,
\end{aligned}
\end{equation}
which can be identified to the time evolution of the degree of orientation given in Eq.~\eqref{cosFT}. This leads to the relations:
\begin{equation}
\begin{cases}
\alpha_{j,j+1}|C_{j+1}C_{j}|& =\frac{A_{0}}{r(1-r)(j+1)^{2}\pi^{2}} \sin[\pi (j+1)r],\\
\phi_{j}-\phi_{j+1}&=-\pi (j+1) r -\frac{\pi}{2}.
\end{cases}
\end{equation}
As for the rectangular signal,  $r$ must be taken irrational. A sawtooth signal can be obtained by considering the Taylor expansion of the triangular response at first order around $r=1$: $1/(1-r)\simeq 1+O(r)$ and $\sin[\pi (J+1)r] \simeq \pi (J+1)r+O(r^{3})$. We arrive at:
\begin{equation}
\begin{cases}
\alpha_{j,j+1}|C_{j+1}C_{j}|& =\frac{A_{0}}{(j+1)\pi},\\
\phi_{j}-\phi_{j+1}&=-\pi (j+1) -\frac{\pi}{2}.
\end{cases}
\end{equation}
Figure~\ref{fig1} displays the time evolution of the degree of orientation when the initial condition is $|\psi_T\rangle$. A reasonable match is achieved between the ideal signal and the response obtained with the target state.
Figure~\ref{fig2} shows the influence of the parameter $j_{\textrm{max}}$ on field-free orientation. As could be expected, the higher $j_{\textrm{max}}$ the better  the signal is. As can be seen in Fig.~\ref{fig2}, a value of $j_{\textrm{max}}=5$ is sufficient to get a nearly perfect sawtooth signal. Figure~\ref{fig2} also illustrates the importance of the correction to the Gibbs-Wilbraham phenomenon.

\subsection{Numerical optimization results}
\begin{figure}[!ht]
	\centering
	\includegraphics[width=1\linewidth]{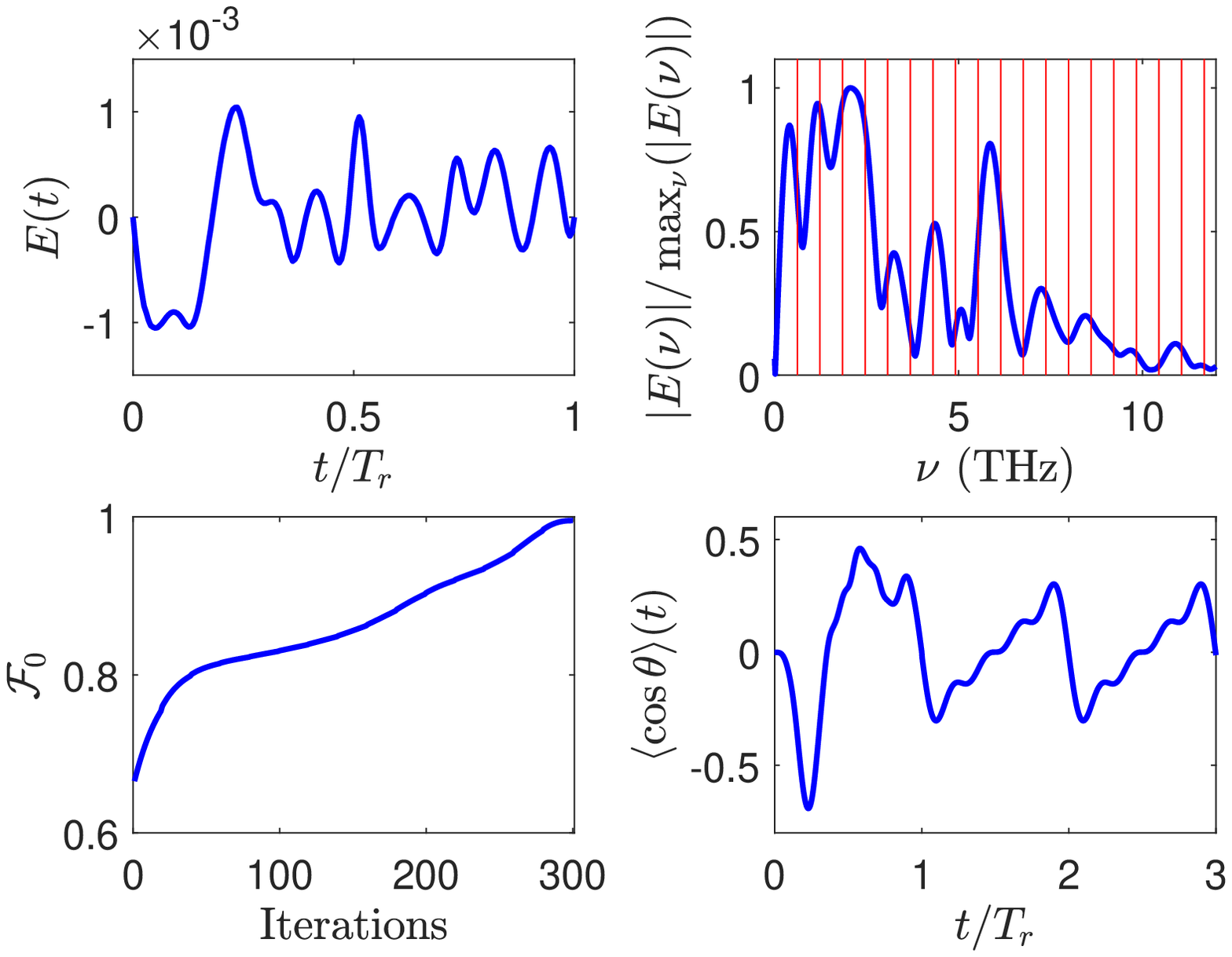}
	\caption{(Color online) (top) Time evolution of the electric field and its normalized Fourier transform. (bottom) Evolution of the figure of merit $\mathcal{F}_0$ as a function of the number of iterations and the corresponding degree of orientation $\langle \cos\theta\rangle$ for the OCS molecule. Numerical parameters are set to $j_{\textrm{max}}=9$, $A_0=1$ and $r=1/\sqrt{2}$.}
	\label{fig3}
\end{figure}

Using optimal control techniques, we have derived a control pulse able to bring the initial state of the system to a target state. The control problem is defined through the figure of merit $\mathcal{F}_0=\Re [\langle \psi(t_0)|\psi_T\rangle ]$ to maximize. The control time $t_0$ is set to $T_r$. A constraint in the design process is used in order to ensure that the field is smoothly switched on and off at the beginning and at the end of the control. We consider a standard GRadient Ascent Pulse Engineering (GRAPE) algorithm, which is a gradient ascent (or descent) based algorithm initially introduced for Nuclear Magnetic Resonance optimal pulse design~\cite{grape}. Various convergence schemes can be used to improve the initial gradient descent approach. In this work, the implementation is based on a L-BFGS second order optimization scheme~\cite{grape2,brysonbook} and the fmincon function of Matlab. The example of the sawtooth signal is represented in Fig.~\ref{fig3}. Very good results are obtained with a final projection $|\langle \psi(t_0)|\psi_T\rangle |^2$ larger than 0.99 after 300 iterations. We observe a monotonic convergence of the algorithm with a final smooth optimal field. The maximum amplitude of the electric field is of the order of $2.5\times 10^8$~Vm$^{-1}$, which is experimentally achievable with the current available THz sources. Figure~\ref{fig3} also shows the Fourier transform of the optimal electric field. The spectral structure of the optimal solution is quite complicated with different peaks close to multiple of $f_r$. Note that simpler control fields could be designed by adding spectral constraints~\cite{lapert:2009}, but with a lower efficiency.

\section{Shaping at non zero temperature}\label{sec4}
\subsection{Introduction}
At a non zero temperature, the design of a target state (a density matrix) corresponding to a desired field-free evolution is more involved. Furthermore, the definition of the target state is not unique since several density matrices can lead to the same result. The problem is therefore difficult to handle with standard optimal control procedures. Instead, we propose to use a Monte Carlo simulated annealing algorithm \cite{ Kirkpatrick671} and we define the figure of merit to maximize the projection of the Fourier coefficients of $\langle\cos\theta\rangle$ onto the ones of the expected time evolution.

The degree of orientation $\langle\cos\theta\rangle (t)$ at time $t\geq t_0~(\tau\geq 0)$ after the extinction of the field can be written as the sum of the individual contributions coming from the different initially populated $|j_0,m_0\rangle$ states weighted by the Boltzmann population~\cite{Stapelfeldt:03,Seideman:05}:
\begin{eqnarray}\label{eq6}
\centering
\displaystyle\langle\cos\theta\rangle(\tau)&=&\sum_{j_0=0}^{\infty}p_{j_0}\sum_{m0_=-j_0}^{m_0=j_0} \langle\psi_{T}^{(j_0,m_0)}(\tau)|\cos\theta|\psi_{T}^{(j_0,m_0)}(\tau)\rangle\nonumber ,\\
\end{eqnarray}
where $p_{j_0}$ is the initial Boltzmann population of the state $j_0$ given by: $p_{j_{0}}=\frac{1}{\mathcal{Z}}e^{-B(j_{0}(j_{0}+1)/k_{B}T}$,
$k_{B}$ being the Boltzmann constant and $\mathcal{Z}$ the partition function of the system. The time evolution of $|\psi_{T}^{(j_0,m_0)}(\tau)\rangle$ can be expressed as:
\begin{equation}\label{timeevolpsi}
|\psi_{T}^{(j_0,m_0)}(\tau)\rangle = \sum_j C_{j}^{j_0,m_0} e^{-iBj(j+1)(\tau)}|j,m_0\rangle .
\end{equation}
The degree of orientation then reads:
\begin{equation}\label{eq6}
\begin{aligned}
\displaystyle\langle\cos\theta\rangle(\tau)=\sum_{j=0}^{\infty} K_j e^{i 2\pi (j+1)f_r\tau}+c.c,
\end{aligned}
\end{equation}
where $K_j$ is the Fourier coefficient of $\langle\cos\theta\rangle(\tau)$ given by:
\begin{equation}\label{eq6}
K_j=\sum_{j_0=0}^{j_{max}}p(j_0)\sum_{m_0=-j_0}^{j_0} \alpha_{j,j+1}^{m_0}C_{j}^{j_0,m_0}C_{j+1}^{*j_0,m_0}.
\end{equation}
Note that only a limited number of rotational levels (up to a given $j_{\textrm{max}}$) are considered in the numerical simulations. The value of $j_{\textrm{max}}$ depends on the temperature, the molecule and the used field strength. The figure of merit to be minimized $\mathcal{F}$ is defined as the distance between the vectors $\vec{K}$ and $\vec{F}$ of Fourier coefficients associated respectively with the degree of orientation $\langle\cos\theta\rangle(\tau)$ and with the targeted time evolution. More precisely, we have:
\begin{equation}\label{eq6}
\mathcal{F}=\frac{||\vec{K}-\vec{F}||}{||\vec{F}||},
\end{equation}
where $\vec{K}=(K_j)_{j=0}^{j_{\textrm{max}}}$.
\subsection{The optimization algorithm}
We use a Monte-Carlo simulated annealing algorithm to design the THz laser field. While genetic algorithms
have been widely exploited in coherent control of molecular alignment and orientation~\cite{Horn,deNalda,RouzeeHertz,Rouzéetheo,Ghafur:09,hertz07,atabek03}, simulated annealing~\cite{Kirkpatrick671} has not been considered, to our knowledge, until now for this purpose. The relative simplicity of the application of such numerical algorithms makes it possible to adapt it straightforwardly to non-standard control problems. However, a relative disadvantage of Monte Carlo algorithms is their low convergence with respect to gradient methods.

We first verify the high efficiency of the algorithm at $T=0$~K where almost perfect projection onto the target state can be reached. Hereafter, we focus on the results obtained at non-zero temperature. For sake of clarity, we describe the different steps of the simulated annealing algorithm used in this work. The field $E(t)$ is taken as a spline interpolation polynomial defined by $N$ points $\{t_i,E_i\}$ where $t_i$ are equally spaced times ranging over one rotational period, $t_i=(i-1)\Delta T$ with $\Delta T=T_r/(N-1)$. The parameters $E_i$ represent the values of the control field, $E_0$ and $E_N$ are set to zero leading to $N-2$ values to estimate. We adapt the standard algorithm to the control problem. The algorithm can be summarized as follows:
\begin{enumerate}
\item Generate $N$ random values of $E_i$ between $-E_0/2$ and $E_0/2$ where $E_0$ is the maximum initial amplitude.
\item Construct the field $E(t)$ as a spline polynomial interpolation through the $N$ points.
\item Evaluate $\langle \cos\theta\rangle$ and $\mathcal{F}$.
\item Set a fictive temperature $T_{MC}$ to an initial value $T_0$ and a maximum number of Monte-carlo iterations $N_{MC}$.
\item Generate a trial field $E_{\textrm{trial}}$ by modifying each value $E_i$ by a small quantity $\Delta E_i$. The displacement $\Delta E_i$ is a random value ranging between $-\Delta E_{\textrm{max}}/2$ and $\Delta E_{\textrm{max}}/2$. The way the maximum displacement $\Delta E_{\textrm{max}}$ amplitude is chosen and updated is discussed later on.
\item Evaluate $\langle \cos\theta_{\textrm{trial}}\rangle$ and $\mathcal{F}_{\textrm{trial}}$ corresponding to $E_{\textrm{trial}}$.
\item  Evaluate the quantity $\Delta \mathcal{F}=\mathcal{F}_{\textrm{trial}}-\mathcal{F}$. If $\Delta \mathcal{F} <0 $ then $E_{\textrm{trial}}$ replaces the reference $E(t)$. In the other case ($\Delta \mathcal{F} >0$), the trial field is accepted with a probability $e^{-\Delta \mathcal{F} / T_{MC}}$. In practice, this means that a random number between 0 and 1 is generated and that the field is changed if this number is lower than $e^{-\Delta \mathcal{F} / T_{MC}}$.
\item The fictive temperature is decreased at each step by a quantity $p T_{MC}$. A parameter $p$ equal to 1/3 is found to be a reasonably good choice.
\item Repeat steps 5-9 until the number of iterations exceeds $N_{MC}$ or until the fictive temperature reaches zero.
\end{enumerate}
The maximum displacement amplitude $E_{\textrm{max}}$ is chosen so that its value decreases as the figure of merit increases. Numerical simulations reveal that good results can be achieved with a function of the form:
\begin{equation}\label{eq6}
\Delta E_{\textrm{max}}=(\mathcal{F}+\kappa e^{(\mathcal{F}-\epsilon)})E_0.
\end{equation}
The parameters $\kappa$ and $\epsilon$ are set to get a reasonable convergence behavior of the algorithm. Table~\ref{tab1} gives the values of the parameters used in the numerical optimizations presented below (unless otherwise specified).
\begin{table}[h!]
\begin{center}
\begin{tabular}{|c|c|c|c|c|c|}
\hline
  $T_0$& $N$ &$\kappa$&$\epsilon$&$N_{MC}$&$E_0[u.a]$\\
  \hline \hline
 $0.1$&$20-60$&$0.08$&$0.01$&$2500$&$6 \times 10^{-5}$\\\hline
\end{tabular}
\caption{Numerical parameters of the Monte-Carlo simulated annealing algorithm.}
\label{tab1}
\end{center}
\end{table}
Note that no constraint on the field area has been introduced in this algorithm. In order to do so, a new term could be added to the figure of merit~\cite{Sugny:14}, but this would be at the detrimental of the projection onto the target. We choose here to do it differently by modifying the step 7 of the algorithm. In the new step 7, we accept the trial field in the case where $\Delta \mathcal{F} >0$ only if the area of the trial field is lower than the area of the current field. We have verified that this method leads to a lower field area with a similar performance when compared to the non constrained algorithm.

\subsection{Numerical results}
\subsubsection{The sawtooth signal}
\begin{figure}[!ht]
	\centering
	\includegraphics[width=1\linewidth]{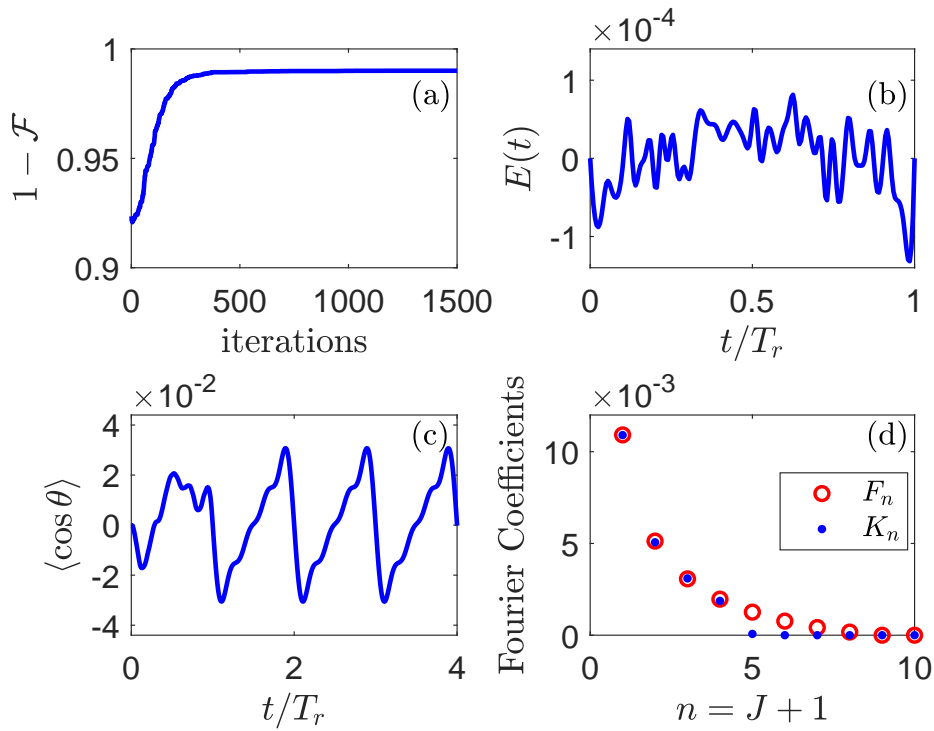}
	\caption{(Color online) Optimization results in the case of a sawtooth signal for the CO molecule at $T=10$~K. Panel (a) represents the evolution of the figure of merit as a function of the number of iterations.
 Panel (b) displays the corresponding optimal electrical field. The time evolution of $\langle \cos\theta\rangle(t)$ is depicted in panel (c). Panel (d) presents a comparison of the modulus of the Fourier coefficients of $\langle \cos\theta\rangle(t)$ and of an ideal sawtooth.}\label{Rampe_T_10K}
\end{figure}

\begin{figure}[!ht]
	\centering
	\includegraphics[width=1\linewidth]{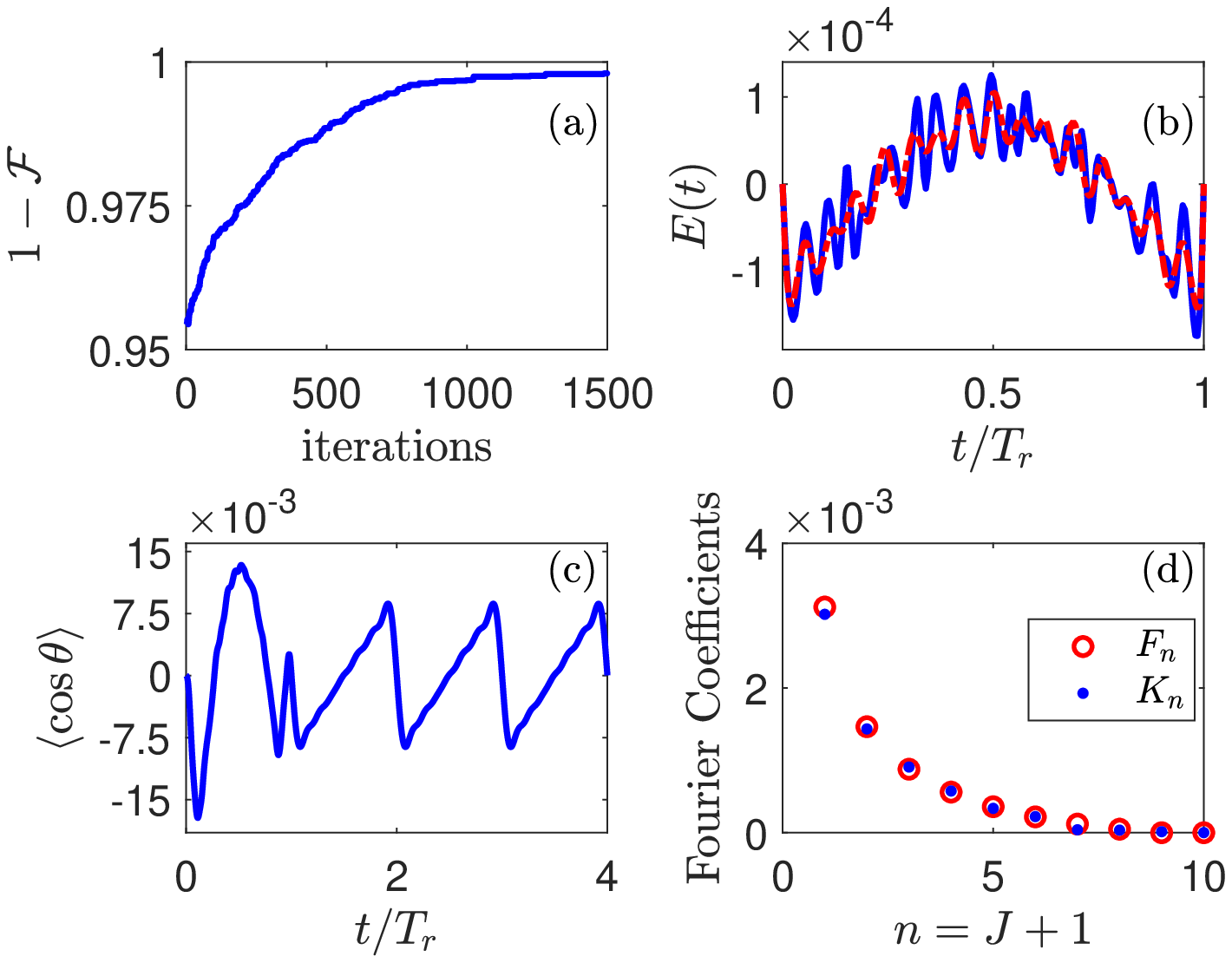}
	\caption{(Color online) Same as Fig.~\ref{Rampe_T_10K} but for $T=30$~K. In panel (b), the dashed red line is the fitted electric field (see the text for details).}
	\label{Rampe_T_30K}
\end{figure}

\begin{figure}[!ht]
	\centering
	\includegraphics[width=1\linewidth]{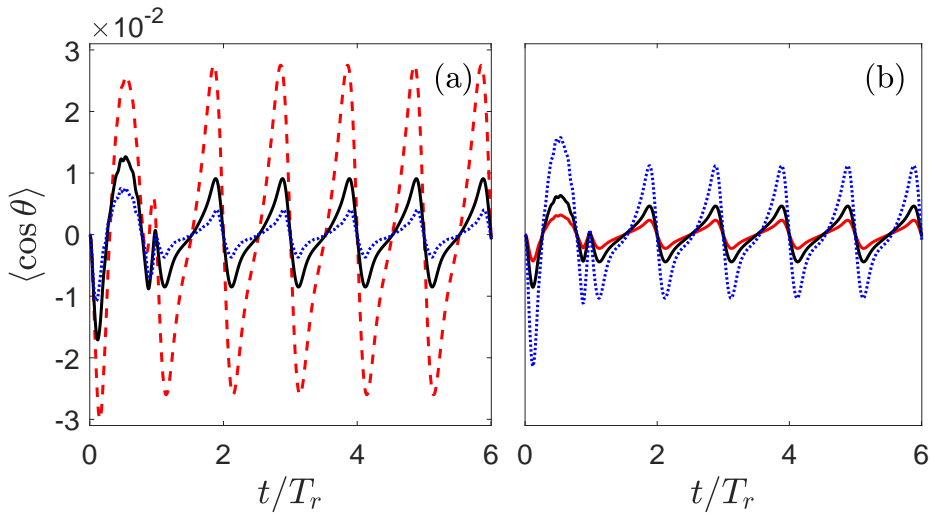}
	 \caption{(Color online) Robustness of the time evolution of $\langle \cos\theta\rangle$ for the CO molecule against temperature effects (a) and amplitude variations of the control field (b). The results in panel (a) were obtained with the fitted electric field $E(t)$ given by the expression \eqref{expressionE} at three different temperatures: 15K (in dashed red), 30K (in solid black line) and 50K (blue dots).
The  numerical parameters used for the field are  displayed  in Tab.~II column (a),the  maximum amplitude is set to $E_m= 1.25\times 10^{-4}$ ~a.u. The results in panel (b) have been obtained at 30K with the same electric field but with a maximum amplitude reduced by a factor $2$ and $4$ (in solid black and red lines respectively) and increased by 25\% (blue dots).}
	\label{Robustesse_rampe}
\end{figure}

We first consider the problem of the generation of a sawtooth signal for the CO molecule at two different temperatures $T=10$~K and  $T=30$~K. The  amplitude $A_0$ of the waveform is respectively taken to be 0.0035 and 0.01. As can be seen in Fig.~\ref{Rampe_T_10K}a and \ref{Rampe_T_30K}a, the algorithm converges smoothly toward a figure of merit close to 0.99 after approximately 1000 iterations. Figures~\ref{Rampe_T_10K}c and \ref{Rampe_T_30K}c show the evolution of the degree of orientation  $\langle\cos\theta\rangle(t)$ obtained after the optimization process. In particular, the response obtained at $T=30$~K is very similar to an ideal sawtooth signal while some oscillations remain at $T=10$~K. This point is confirmed by Fig.~\ref{Rampe_T_10K}d and \ref{Rampe_T_30K}d where the Fourier coefficients of the targeted sawtooth signal and of $\langle\cos\theta\rangle(t)$ are compared at temperatures $10$~K and $30$~K, respectively. We observe that at $T=10$~K, a good agreement is obtained  only for the low $j$ coefficients ($j\leq 4$), the high $j$ coefficients are not well reproduced. This is mainly due to the fact that a weak electrical field predominantly induces transitions between adjacent rotational levels and that only 3-4 rotational levels are significantly initially  populated at 10~K. At $T=30$~K, more than 10 rotational levels are significantly populated, which explains the good matching of the high Fourier coefficients. The  optimal electric fields at the two temperatures (see Fig.~\ref{Rampe_T_30K}b and Fig.~\ref{Rampe_T_10K}b) present similar features such as a bell-shaped envelope, the carrier being a fast oscillating signal. The field shape has a simple form at $T=30$~K and is very close to a cosine envelop with a period of the order of $T_r/2$. We have performed a least square fit of the electric field obtained at $T=30$~K with a function of the form:
\begin{eqnarray}\label{expressionE}
& E(t)= E_m \Big(E_1\sin(2\pi f_1 t+\phi_1)+E_2\sin(2\pi f_2 t+\phi_2)\nonumber \\
  & +E_3\sin(2\pi f_3t+\phi_3)+E_0\Big)\Pi_{\sigma_1,\sigma_2}(t)
\end{eqnarray}
where $E_m$ is the maximum amplitude of the field which is set to $ 1.25\times 10^{-4}$~a.u. as in Fig.~\ref{Rampe_T_30K}. $\Pi_{\sigma_1,\sigma_2}$ is a window function of period $T_r$ with a finite rise (resp. fall) time $\sigma_1$ (resp. $\sigma_2$) defined as follows:
\begin{equation}\label{window}
\Pi_{\sigma_1,\sigma_2}(t)=
\begin{cases}
1-e^{-\frac{t}{\sigma_1}}, & 0<t\leq\frac{T_{r}}{2},\\
1-e^{-\frac{T_r-t}{\sigma_2}}, &  \frac{T_{r}}{2}<t<T_{r}.
\end{cases}
\end{equation}
The different parameters ($E_i$,$f_i$,$\phi_i$,$\sigma_1$,$\sigma_2$) are determined by a least square fit and are given in Tab.~II. The field consists mainly in a superposition of a strong DC component combined with a sine function  of frequency $0.56 f_r$ and a weaker sine component of frequency $15.5 f_r$ in phase opposition with the previous signal. The signal is gated by a rectangular window of  width $T_r$ with a rising time of $0.063 T_r$ and a fall time of $0.036 T_r$. As can be seen in Fig.~\ref{Rampe_T_30K}, the fitted electric field reproduces globally the behavior of the optimal field. We investigate in Fig.~\ref{Robustesse_rampe} the robustness of the derived analytical solution against temperature effects and amplitude variations of the control field. It is remarkable that a sawtooth like signal is still observed at $T=15$~K and 50~K. When the maximum amplitude $E_m$ is decreased, no significant shape variation is observed as can be seen in Fig~\ref{Robustesse_rampe}b  where the amplitude $E_m$ is reduced by factors 2 and 4. However  distortions occur for an amplitude increase of 25\%, the targeted shape being completely lost for an increase of 50\%.

\begin{table}\label{table2}
\begin{center}
\begin{tabular}{|c|c|c|}
\hline
  Parameter&Case(a)&Case(b)\\
  \hline
  \hline
  $E_m[a.u]$&$1.25\times 10^{-4}$&$1.89\times 10^{-4}$\\
  $E_0$&-0.7876&-0.0228\\
  $E_1$&1.36&0.7989\\
  $E_2$&0.1679&0.1138\\
  $E_3$&0.1259&0.0307\\
  $f_{1}/f_r$&0.56&1.0064\\
  $f_{2}/f_r$&15.54&8.3\\
  $f_{3}/f_r$&11.137&3.31\\
  $\phi_1[\pi]$&-0.0634&0.9631\\
  $\phi_2[\pi]$&1.082&0.5906\\
  $\phi_3[\pi]$&0.036&-0.38\\
  $\sigma_{1}/T_r$&0.063&0.0504\\
  $\sigma_{2}/T_r$&0.036&5 $10^{-2}$\\
  \hline
\end{tabular}
\end{center}
\caption{Parameters obtained by a least square fitting of the electrical field with Eq.~\eqref{expressionE}. The cases (a) and (b) correspond respectively to a sawtooth signal at $T=30$~K (see Fig.~\ref{Rampe_T_30K}) and to a rectangular signal at $T=30$~K (see Fig.~\ref{creneau_T_30K}).}
\end{table}

\subsubsection{The rectangular signal}
\begin{figure}[!ht]
	\centering
	\includegraphics[width=1\linewidth]{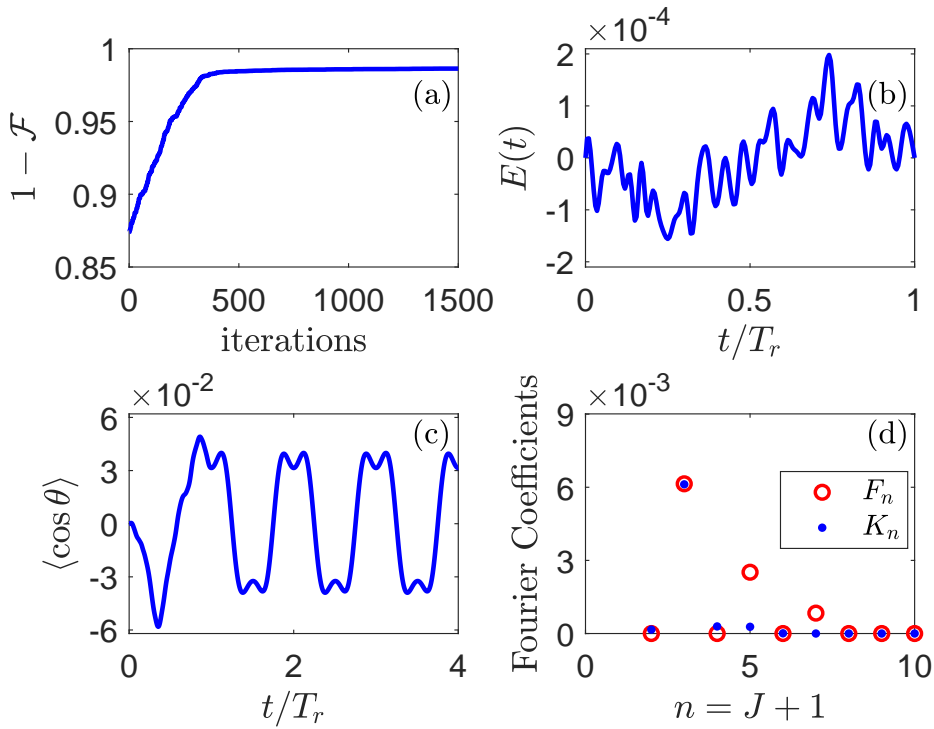}
	\caption{(Color online) Optimization results in the case of a rectangular signal for the CO molecule at $T=10$~K.
 Panel (a) represents the evolution of the figure of merit as a function of the number of iterations. Panels (b) and (c) show respectively the optimal electrical field and the corresponding time evolution of the degree of orientation. A comparison between the modulus of the Fourier coefficients of $\langle \cos\theta\rangle(t)$ and of an ideal rectangular signal is presented in panel (d).}
	\label{creneau_T_10K}
\end{figure}

\begin{figure}[!ht]
	\centering
	\includegraphics[width=1\linewidth]{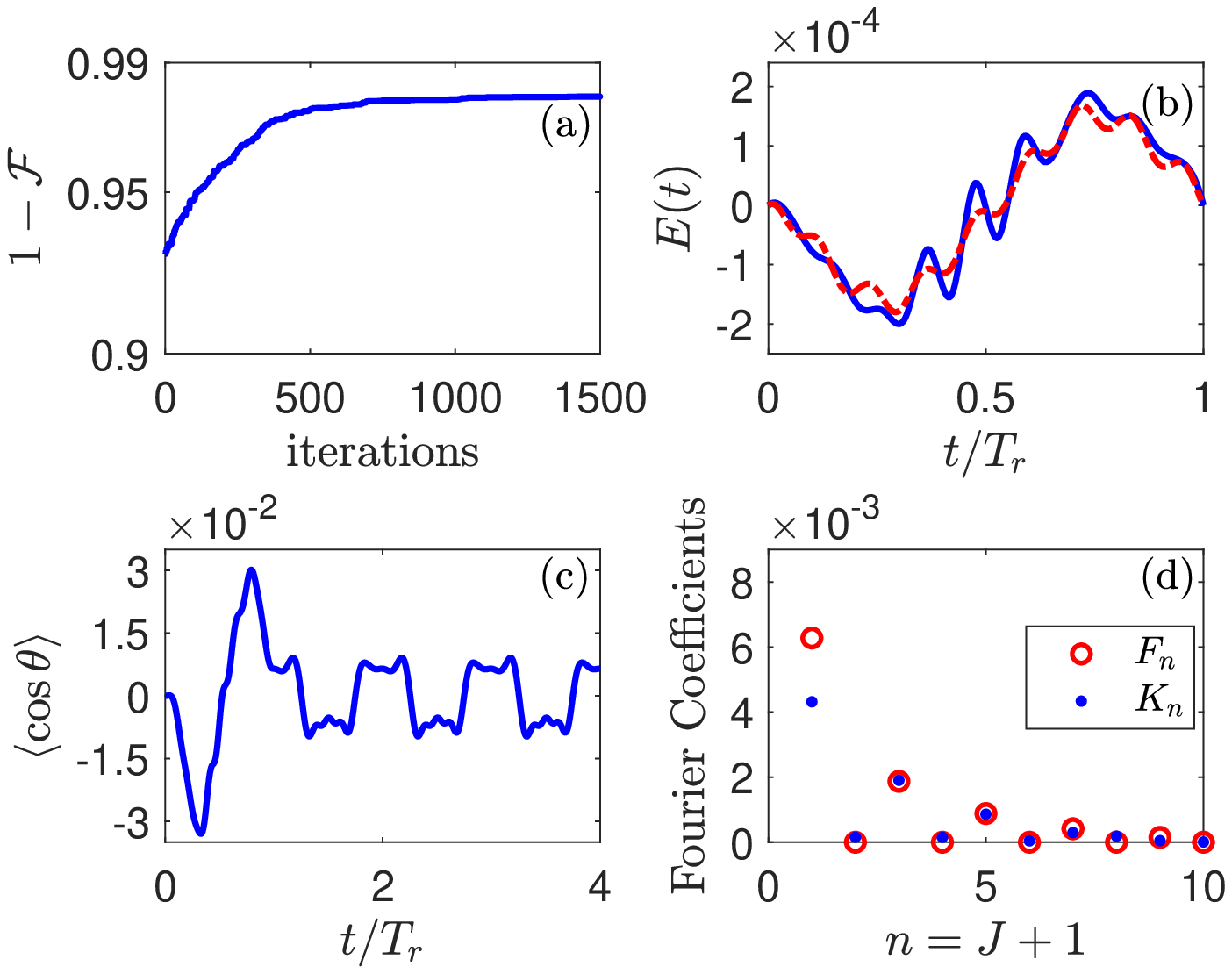}
	\caption{(Color online) Same as Fig.~\ref{creneau_T_10K} but for $T=30$~K. In panel (b), the dashed red line is the fitted electric field (see the text for details)}
	\label{creneau_T_30K}
\end{figure}

We address in this paragraph the case of a rectangular signal. We recall that at $T=0$~K, periodic rectangular signals with a rational $r$ parameter cannot be designed due to the cancellation of some Fourier coefficients. At a temperature different from zero, such cases are in principle possible because the zero Fourier coefficients can be obtained by an interference effect coming from the responses of different initially populated rotational $j_0$ levels. Thus no particular condition on the $r$ parameter of the rectangular signal is required. We have taken $r=0.5$ as an illustrative example. The results for  the CO molecule at two different temperatures  $T=10$~K and $T=30$~K and for an amplitude $A_0$ of the rectangular waveform equal to 0.035 and 0.01 respectively are presented in Fig.~\ref{creneau_T_10K} and \ref{creneau_T_30K}. In both cases, the convergence of the algorithm is reached after approximately 500 iterations. As for a sawtooth signal, the  degree of orientation at $T=30$~K presents less oscillations than at $T=10$~K. It is worth noting that at $T=30$~K a small discrepancy occurs for the first Fourier coefficients due probably to the initial Boltzmann distribution.

\begin{figure}[!ht]
	\centering
	\includegraphics[width=1\linewidth]{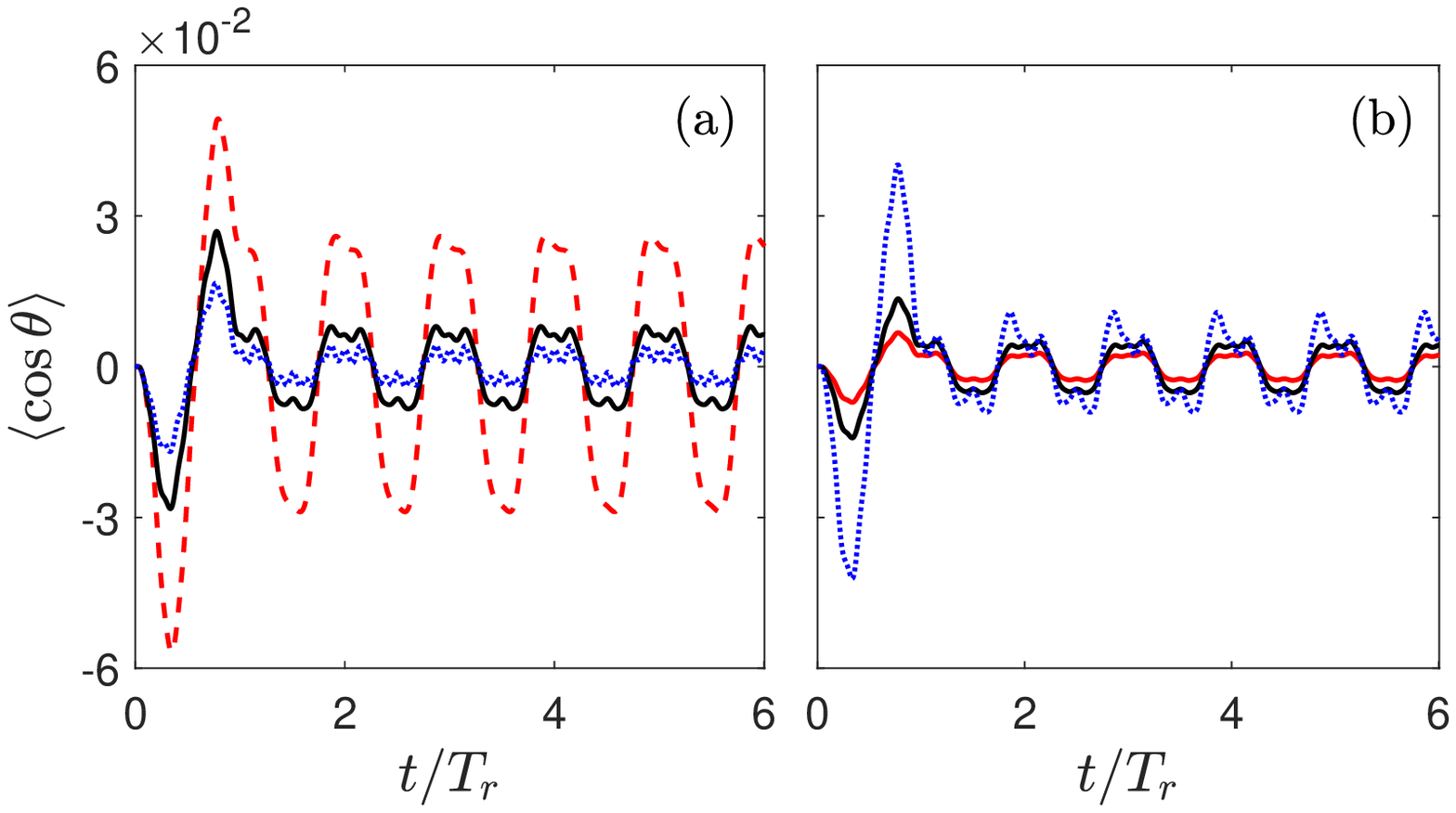}
	\caption{(Color online) Same as Fig.~\ref{Robustesse_rampe} but for the fitted electric field with the parameters given in the column (b) of Tab.~II.}
	\label{Robustesse_crenau}
\end{figure}

We have performed a least square fit of the field according to Eq.~\eqref{expressionE}. The corresponding parameters are displayed in the right column of Tab.~II. As can be seen in Tab.~II, the field consists mainly in the superposition of two cosine functions with frequencies equal to $f_r$ and $8.3 f_r$. The two components of the signal are in phase quadrature. We have also investigated the robustness against temperature and amplitude variations. The results are presented in Fig.~\ref{Robustesse_crenau}. We observe that the overall shape of the degree of orientation is preserved at 15 and 50~K. At larger temperature, rapid oscillations occur because high-frequency Fourier coefficients are not well reproduced. At lower temperature, the signal becomes sinusoidal. As it was observed for the sawtooth case, no significant shape variation occurs when the amplitude of the field is decreased. However, when the amplitude is increased, a significant distortion can be seen for a variation larger than 25~\%.

\color{black}

\section{Conclusion}\label{sec5}
We have investigated in this work the extend to which the time evolution of field-free molecular orientation can be shaped. We have shown that the degree of orientation can be steered along a predefined periodic signal with a time-integrated zero area. Rectangular, sawtooth and triangular functions are taken as examples. At zero temperature, we have shown how to design a target state corresponding to the desired signal. The target state can reached by using optimal control procedures with a very good efficiency. At non-zero temperature, the target state is not uniquely defined and it is easier to consider a figure of merit corresponding to the normalized distance between the Fourier coefficients of the degree of orientation and of the targeted signal. We have used a specially designed Monte-Carlo simulated annealing algorithm for maximizing this figure of merit. The optimization results lead to a good agreement of the designed orientation with the expected signal, which can be slightly better at high temperatures. In the different cases, an analytical expression for the electric field can be derived as a superposition of sinusoidal functions with different phases and gated by a temporal rectangular window with finite rising and fall times. The derived solution was found to be robust against variations of the amplitude of the field and temperature effects. The observed robustness is very interesting from the experimental point of view since it makes the optimal electric field insensitive to thermal fluctuations and to spatial inhomogeneities of the field.
\begin{figure}[!ht]
	\centering
	\includegraphics[width=1\linewidth]{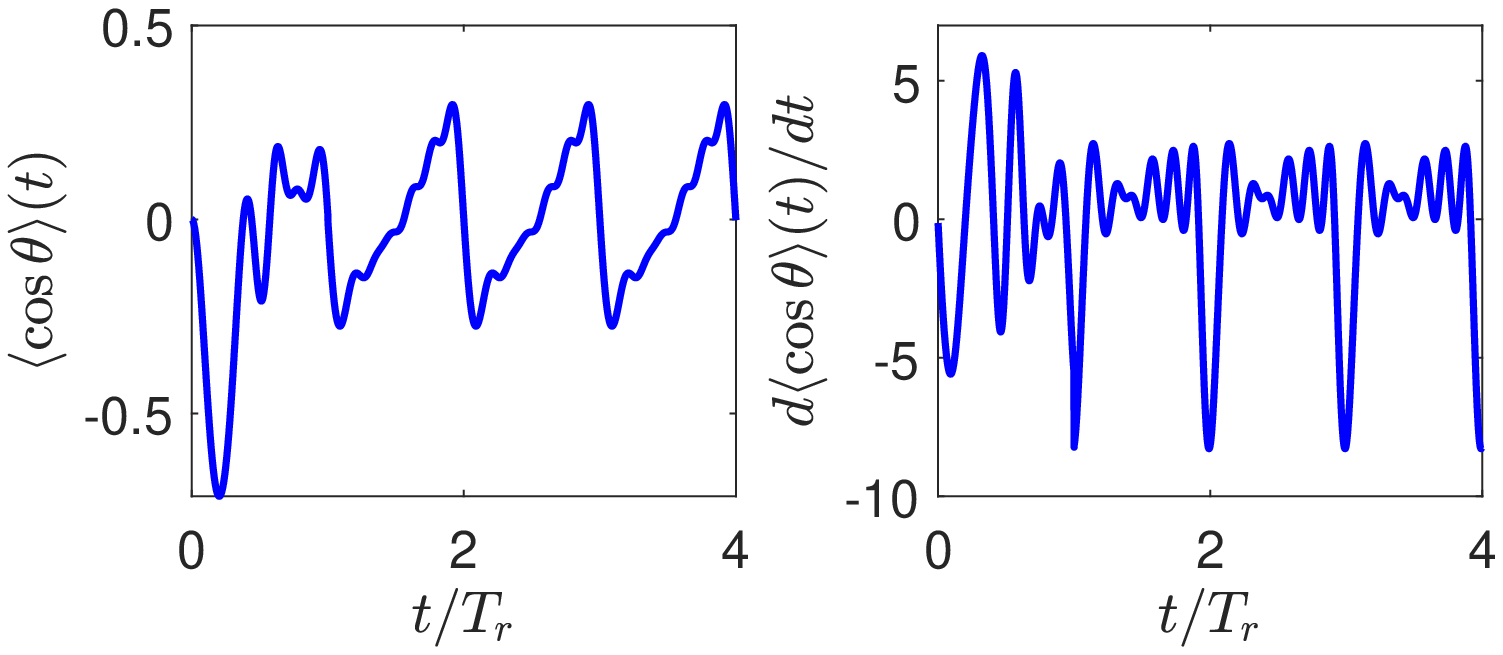}
	\caption{(Color online) Time evolution of the degree of orientation (left) and of its time derivative (right) at $T=0$~K. The first rotational period corresponds to the application of the electric field, the three others to field-free dynamics. The time derivative function is expressed in arbitrary units.}
	\label{peigne}
\end{figure}
The shaping of THz pulses has known in recent years an impressive experimental development (see e.g.~\cite{amico:2009,vidal:2014,kohli,stepanov,gingras} to mention a few). These studies show that the shape of the generated THz waveform can be optimized to some extent. The central frequency can be tuned and the width of the spectrum can be modified. In Sec.~\ref{sec4}, we have shown that the optimal control field can be approximated by the sum of two or three sinusoidal functions with a specific amplitude. Such fields could be generated experimentally in a near future in view of recent experimental progress. In addition, the robustness against temperature effects and field variations of the optimized field is a key point to apply such pulses in different experimental conditions and to achieve a noticeable degree of orientation. We have also verified that the first two higher moments $\langle\cos^3\theta\rangle$ and $\langle\cos^5\theta\rangle$ have a similar time evolution as $\langle\cos\theta\rangle$. This behavior could be interesting in the case of specific signals such as a laser induced ionization which exhibit a non linear behavior with respect to $\cos\theta$~\cite{ionization}. Finally, we point out that the shaping of the time evolution of the orientation signal should be also possible with a spectrally shaped two-color laser pulse~\cite{vrakking:1997,dion:1999,Tehini:08,Znakovskaya:09}. This issue which goes beyond the scope of this study is an interesting generalization of the results presented in this paper.

The potential applications of this work may be  found in the temporal or spatial control of ionization and birefringence~\cite{Stapelfeldt:03,Seideman:05} or in the generation of THz clocks for telecommunications and metrology (when a two-color laser pulse is used to shape the rotational dynamics). In order to explore this latter application, we have plotted in Fig.~\ref{peigne} the time derivative of the degree of orientation produced by the control processes at zero temperature. As shown in \cite{Nelson:11,Babilotte:16}, this function is proportional to the THz field emitted by the sample and is thus directly measurable. As could be expected, this function is very close to a Dirac comb for a sawtooth signal. The peaks have a width of the order of $0.15\times T_r$. The shaping of field-free orientation can thus be viewed as a way to produce a THz Dirac comb, which could be very useful for THz clocks.

\end{document}